# Optimized hydrogen sensing properties of nanocomposite NiO:Au thin films grown by dual Pulsed Laser Deposition

I. Fasaki [1], M. Kandyla[1], M. G. Tsoutsouva[1], M. Kompitsas[1*]

[1]*National Hellenic Research Foundation, Theoretical and Physical Chemistry Institute, 48 Vasileos Konstantinou Ave., 11635 Athens, Greece*

**Abstract**

Nanocomposite NiO:Au thin films, formed by gold nanoparticles embedded in a nickel oxide matrix, have been grown by reactive pulsed laser deposition (R-PLD). Two actively synchronized nanosecond laser sources, a KrF excimer laser (248 nm) and a Nd:YAG laser (355 nm), were used for the simultaneous ablation of nickel and gold targets in oxygen ambient. The morphology, composition, and optical properties of the obtained nanocomposites were investigated and were found to correlate with the concentration of Au nanoparticles. Further, the NiO:Au nanocomposites have been tested as hydrogen sensors. Embedding Au nanoparticles into the NiO film matrix reduced the sensors operating temperature and improved their performance by orders of magnitude.

* Corresponding author:
M. Kompitsas
National Hellenic Research Foundation
48 Vas. Konstantinou Ave
11635 Athens, Greece
Tel. +302107273834
Fax +302107273794
E-mail: mcomp@eie.gr
URL: www.laser-applications.eu

## 1. Introduction

Hydrogen sensing is important for many industrial applications because hydrogen becomes explosive in air with a lowest explosion limit (LEL) of 4% (40,000 ppm). Hydrogen used as fuel is a major concern for fire detectors: traditionally, fire alarms use a smoke detector or a heat detector. In the event of fire, however, various combustible gases are also produced and those gases, especially hydrogen, diffuse more rapidly than smoke or heat does. Therefore, 1/10 of the lowest explosion limit for each gas is used as the alarming level for gas sensors. The challenges of the gas sensor technology today include improving the performance of the devices, realizing room temperature sensors, and developing micro and nanosensors, among others.

Recent studies have shown that NiO thin films can be used successfully as sensing materials in gas and humidity detection devices [1,2]. NiO, one of the very few known p-type metal-oxide semiconductors [3], is a dielectric, antiferromagnetic, electrochromic [4,5], and catalytic [6] material with excellent chemical stability [7] and good gas sensing properties [8,9]. NiO thin films have been prepared in the past by several methods, such as sputtering [9], chemical spray pyrolysis [10], and Pulsed Laser Deposition (PLD) [4,5,11,12].

Thin films containing noble metal nanoparticles have become an intensive area of research due to their interesting functional properties that find applications to the electronic, glass, sensing, catalysis, semiconducting, and computing industries [13]. The goal of the research in this area is to produce nanoparticles of a specific size, shape, and concentration, either on the film surface or embedded within the host matrix or a combination of both. In particular, nanocomposite metallic oxide thin films with Au nanoparticles find extensive applications in sensor devices, including among others: electrochemical $SnO_2$:Au sensors for CO with improved performance [14]; biochemical sensors with enhanced response [15]; optical detection of CO based on absorbance changes of NiO:Au compounds [16]; gas detection selectivity between CO and $H_2$, by means of optical sensing based on CuO:Au compounds [17], and ozone detection by conductometric $SnO_2$:Au gas sensors [18].

Nanocomposite NiO:Au has been developed previously by pyrolysis [16,17]. The Au nanoparticles had a diameter of a few tens of nm and induced a catalytic enhancement of the optical detection of CO. A new chemical process for nanocomposite thin-film growth has been proposed and developed recently [19]. Metallic (*e.g.*, Au, Ag, Pt) or narrow-bandgap semiconducting (*e.g.*, CuO) nanoparticles, embedded between $SnO_2$ crystallites, served as nanoelectrodes, reducing the resistivity of the electrochemical solid-state gas sensor and leading to a sensing device with improved performance. In [19], the nanoparticles have been deposited either below ("bottom layout") or on the surface ("top layout") of the functional $SnO_2$ film. In the present paper, we report on a new type of NiO:Au nanocomposite thin films, where Au nanoparticles are stochastically embedded inside the NiO film matrix ("intermediate layout"). This growth technique has been previously employed [20] as an extension of the traditional PLD technique: the setup consists of a synchronized dual-laser, dual-target system that allows the control of the concentration of any dopant into any thin-film matrix. This is achieved by adjusting the fluence of the laser beam incident on the dopant target during film growth. In this way, the properties of the nanocomposite material may be efficiently optimized for specific applications.

In the present study, special emphasis has been placed on correlating the fluence of the laser beam incident on the Au target, with the NiO:Au nanostructure properties, such as Au concentration, film surface morphology, and optical properties. Additionally, the sensing response of the NiO:Au nanocomposites to hydrogen has been systematically investigated. Hydrogen detection was achieved at operating temperatures much lower than those found in the literature [8], therefore the energy consumption of the sensor was reduced without sacrificing its detection efficiency. Ultimately, the detectable hydrogen concentration in air was reduced to 5 ppm for an operating temperature of 135 ºC.

## 2. Materials and methods

The experimental method has been described in a previous publication [21]. Briefly, a two-laser, two-target Reactive Pulsed Laser Deposition (R-PLD) technique was employed (Figure 1). The first laser was a 248-nm KrF excimer laser and was used for the ablation of the Ni metallic target. The second, 355 nm Nd:YAG laser was used for the ablation of the Au target. Both lasers had $\tau \sim 10$ nsec pulse duration and were actively synchronized at 10 Hz pulse repetition rate. The KrF laser fluence was constant at 5.5 J/cm$^2$. The Nd:YAG laser fluence was set at 1.3, 2.4, and 4.3 J/cm$^2$.

The Ni and Au targets were placed in a high vacuum chamber. The depositions occurred at 40 Pa dynamic pressure of oxygen. Quartz and Si/SiO$_2$ (190 nm SiO$_2$ layer) were used as substrates, positioned 40 mm away from the targets. The two target ablation plumes intersected each other on the substrate surface. The substrates were heated at 400 °C during deposition, which typically lasted for 1.5 hour.

The surface morphology of the NiO:Au compounds was investigated with the aid of a JEOL JSM 6380-LV Scanning Electron Microscope (SEM), equipped with an Energy Dispersive Spectrometer (EDS). Rutherford Backscattering Spectroscopy (RBS) was performed with the aid of a tandem ion accelerator at *NCSR 'Demokritos'* [12] and revealed both the thickness and the Au concentration of the NiO:Au nanocomposites. Optical absorption spectra were recorded with a Perkin Elmer Lambda 19 spectrophotometer in the 300-1300 nm range.

Hydrogen sensing tests were performed in an aluminum vacuum chamber filled with dry air at atmospheric pressure. The samples were resistively heated and the hydrogen concentration was calculated based on the partial pressures by an MKS Baratron gauge. At a constant bias voltage of 1 V, the current through the samples was recorded by a Keithley 485 picoammeter. Current changes were monitored in real time and the sensor response $S$,

$$S = \frac{R_g - R_o}{R_o} \tag{1}$$

was calculated for various hydrogen concentrations and operating temperatures. Here, $R_g$ and $R_0$ are the sensor resistance with and without hydrogen, respectively.

Initially, all tests were performed under static gas pressure conditions. Pure hydrogen was obtained directly from a bottle and the least partial pressure that could be measured was limited by the accuracy of the Baratron gauge. In a later stage, we employed two computer-controlled mass flow meters (Bronkhorst High-Tech) that allowed us to perform sensing tests under dynamic pressure (flow) conditions. In order to try out smaller concentrations, hydrogen was mixed with dry nitrogen in a premixing chamber, thus achieving dilution factors below $10^{-2}$. With the aid of this improved setup, we were able to detect hydrogen at concentrations down to the level of a few ppm.

## 3. Results and Discussion

### 3.1. Surface morphology and composition

In Figure 2, SEM and EDS data from a reference, pure NiO thin-film sample, grown on a Si/SiO$_2$ substrate by PLD are presented. The SEM image in Fig. 2a shows that the reference thin film is smooth and the surface is dominated by spherical droplets of various sizes, a typical observation for PLD-grown thin films. The EDS spectrum in Fig. 2b results from the "film background" (area outside droplets). The EDS spectrum in Fig. 2c results from the droplet indicated in Fig. 2a. The increase of the Ni signal in Fig. 2c, compared to the Ni signal in Fig. 2b, indicates that such droplets consist of metallic Ni target material that has not been fully oxidized in the laser-produced plasma. The largest diameter of these droplets is estimated to be on the order of 1 μm.

EDS spectra from the "film background" of NiO:Au nanocomposite thin films, deposited with 1.3 J/cm$^2$ (Fig. 3a) and with 4.3 J/cm$^2$ (Fig. 3b) fluence of the Au ablating laser, indicate the increased Au amount inside the film host with increased energy density of the second laser, as expected. Figure 3c shows an SEM image of a NiO:Au compound film, deposited with fluence 4.3 J/cm$^2$ of the Au ablating laser, and Figure 3d the EDS spectrum of a droplet (estimated size of ~ 1 μm) on the surface of the film. The EDS spectrum indicates that the droplet consists mainly of Au. From the above observations, we conclude that Au is both incorporated inside the NiO host and randomly distributed on the NiO surface. XRD spectra taken on the NiO:Au nanocomposite films, show that NiO and Au crystallize during laser deposition, with average crystallite sizes of 46 and 22 nm, respectively [21].

To investigate the film composition and to seek further evidence for the presence of Au in the nanocomposite films, we employed the RBS technique, which we have employed in PLD-grown NiO films in the past proving they are stoichiometric [12]. A recorded RBS spectrum, obtained for the NiO:Au sample deposited with 1.3 J/cm$^2$ laser fluence, along with the corresponding SIMNRA simulation [22], is shown in Fig. 4a. Oxygen nuclei were used as projectiles, therefore only the heavier Ni and Au elements have been recorded. The two fitted parameters in the simulation were the amount of Au atoms (in percent, relative to the total amount of Au and Ni atoms) and the film thickness. The fitted parameters for each sample are given in Table 1, with respect to the fluence of the Au ablating laser**.** In Fig. 4b the percentage concentration values of Au are plotted against the laser fluence, showing a linear relation for the employed laser fluence range. This curve can be used to tailor NiO:Au nanocomposite films for specific applications.

### 3.2. *Optical Measurements*

Noble metal nanoparticles interact strongly with light through the resonant excitation of collective oscillations of their conduction electrons (Surface Plasmon Resonance, SPR). The SPR band characteristics are related to the nanoparticles size [13,23], shape [24], and volume fraction [25], and the dielectric properties of the host matrix, in which the nanoparticles are embedded [26,27]. We have recorded the absorbance spectra of the deposited nanocomposite thin films in the 300-1300 nm spectral range and the result is shown in Fig. 5a. For comparison purposes, we have also recorded the absorbance spectra of the quartz substrate as well as the reference NiO thin film. The broad absorption peak of the NiO:Au samples, centered approximately at 560 nm, is attributed to the SPR of the Au nanoparticles. This peak is not very sharp, indicating a broad distribution of the size of the Au nanoparticles. As a general rule, the intensity of the SPR band increases with the number of nanoparticles [28]. This fact correlates well with the observation in Fig. 5a, where the increase of the fluence of the Au ablating laser results to an increasing intensity of the absorption peak. No significant shift to longer wavelengths (red shift) with the laser fluence is observed and this means that the size of the nanoparticles remains practically unchanged [27,29].

The measured absorbance, $A$, was converted into absorption coefficient, $\alpha$, using the relation

$$\alpha = [A \times \ln 10]/d \qquad (2)$$

where $d$ is the thickness of the films. The Tauc plot in Fig. 5b shows the dependence of $(h\nu\alpha)^2$ on the photon energy, $h\nu$. The bandgap values of the films, $E_g$, are estimated from the dependence $(h\nu\alpha)^2 \sim (E-E_g)$, using the interception values of the tangents to the linear part of the $(h\nu\alpha)^2$ plots with the photon energy axis. The bandgap decreases monotonically with the increase of the fluence of the Au ablating laser due to the incorporation of Au nanoparticles into the NiO film matrix. This effect may be useful for various optoelectronic applications of NiO and other metal-oxide materials, as it has also been observed with $TiO_2$:Au nanocomposites grown by a similar dual-laser, dual-target PLD technique [20], as well as $TiO_2$:Au [26,30] and $WO_3$:Au films [27] deposited by sol-gel techniques.

### 3.3. Hydrogen sensing

Pulsed laser-deposited NiO is a p-type semiconductor and the charge carriers are the holes in the valence band. Hydrogen sensing manifests itself in the reduction of the film conductivity, as it was shown in [12]. In the present work, the sensor response has been calculated according to Eq. (1) and plotted versus time. All samples were tested at temperatures 84, 96, 117, 128, 142, and 168 °C and hydrogen concentrations of 1000, 2000, 5000 και 10000 ppm under static gas pressure conditions. Figure 6 shows a typical sensor response versus time of (a) NiO and (b) NiO:Au (with 44.4% Au) for various hydrogen concentrations at 168 °C operating temperature. It is evident that the incorporation of Au nanoparticles both increases the response *S* of the sensor as well as slightly reduces the response time (defined as the time interval between 10% and 90% of the total signal change), *e.g.*, from 9.5 to 3.5 minutes for 5000 ppm hydrogen.

Figure 7 summarizes the response of NiO and all NiO:Au nanocomposites with respect to hydrogen concentration and operating temperature. The reference sample NiO (Fig. 7a) and the sample with 15.6% Au (Fig. 7b) responded only at temperatures 141 and 167 °C. In Fig. 7b the response increased remarkably for the operating temperature 167 °C. The incorporation of more Au nanoparticles into the NiO matrix, Figs. 7c and 7d, allowed the reduction of the operating temperature without sacrificing the sensor response. In particular, the NiO:Au sample with 44.4% Au (Fig. 7d) responded to hydrogen even at 84 °C. While for all other samples the response increases with the operating temperature, the response of the sample with 44.4% Au reaches a maximum at 128 °C and then drops to lower values as the temperature is further increased.

The mechanism for hydrogen detection in air for the p-type NiO semiconductor is the following [8]: Initially, atmospheric oxygen is adsorbed on the NiO grains surface as $O^-$ or $O_2^-$ by removing electrons from the NiO grain. Thus the number of holes (majority charge carriers) increases and the overall film conductivity increases as well. When $H_2$ reacts with the adsorbed oxygen on the NiO grains surface, the reaction yields $H_2O$ vapor and the delocalized electrons recombine with the NiO holes and the film conductivity decreases.

The embedded Au nanoparticles contribute both to the reaction between hydrogen and oxygen, acting as catalysts, as well as to charge transport in the device, acting as nanoelectrodes, depending on their size. From the structural and optical data presented above, we know the Au nanoparticles have an average size of 22 nm with a rather broad size distribution, while the average NiO crystallite size is 46 nm. The larger Au nanoparticles are dispersed between NiO grains as nanoelectrodes and form new electrical paths, decreasing the film resistivity. Indeed, we have observed the film resistivity drops exponentially with the concentration of Au nanoparticles [21]. Therefore, the electric current of the device which is recorded as the sensor signal, increases, providing high signal-to-noise ratio. Thus, very small changes in the current, arising from the presence of ppm-level concentrations of hydrogen, can be resolved in our measuring apparatus, leading to an improvement of the sensor performance. Similarly, the decrease of the film resistivity due to the presence of embedded Au nanoelectrodes allows for a lower working temperature while maintaining electric current values acceptable for electrochemical sensing. Low film resistivity is important for portable sensors, because it reduces the power required to heat the device to operating temperatures.

Smaller Au nanoparticles are dispersed on the surface of NiO nanocrystallites and function as catalysts: they improve the sensor performance by a *chemical* and by an *electronic* interaction [31,32]. According to the *chemical* "spill over" process, an Au nanoparticle dissociates both the $O_2$ (in the initial adsorption process) and the $H_2$ (during the sensing process) molecules and the atoms spill over onto the surface of the supporting semiconductor, leading to a faster reaction of H and $O^-$ ions [18]. In the second *electronic* process, we neglect the oxygen atoms adsorbed on the semiconductor and consider the adsorbed oxygen on the Au catalyst itself: this is the case when the specific growth process has led to a large dispersion of the catalyst Au nanoparticles on the NiO grains. First, it results in a large increase of the effective sensor surface. Secondly, the adsorbed oxygen removes electrons from the catalyst and the latter removes electrons from the supporting semiconductor. This increases the number of holes and thus increases the sensor conductivity. Therefore, when $H_2$ removes the adsorbed oxygen on the catalyst, the overall sensor conductivity decreases again. The combination of the chemical and electronic mechanisms results in larger changes in the recorded signal in the presence of Au nanoparticles. The question which mechanism dominates may be answered in a separate study, as it is beyond the scope of this work.

Figure 7d indicates that the relation between the sensor sensitivity and the operating temperature *T* can be used for gas selectivity in semiconductor gas sensors. It was shown [31] that for nanocomposite thin film sensors the sensitivity goes through a maximum value for an operating temperature $T_m$ and then drops when the temperature is further increased. The temperature $T_m$ depends on the sampling gas and the catalyst type (usually Pt and Pd). This was explained by the ''burning effect'' [32], namely that for temperatures above $T_m$ the sampling gas oxidizes before it is detected on the sensor surface. Among all reducing gases, hydrogen showed the lowest $T_m$ value [31]. Figure 7d shows that for the sample with 44.4% Au, $T_m$ for hydrogen is around 128 $^o$C, a fact that can be used for gas selectivity *e.g.,* by a sensor array where each sensor operates at a different temperature, optimized for a specific gas.

Figure 8 shows the response of a NiO:Au sensor operating at 135$^o$ C under flow conditions. The detected concentration range is 60 – 5 ppm hydrogen in air and the NiO:Au sample contains 44.4% Au. The response time of the sensor increases slightly with reducing hydrogen concentration as expected and ranges from ca. 11 min for 60 ppm to 14 min for 5 ppm hydrogen concentration. There was a time interval of several months between the static measurements presented in Fig. 6 and the development of the setup for the flow measurements presented in Fig. 8. However, there was no degradation in sensor performance, demonstrating the stability of the device.

## 4.    Conclusions

Applying a versatile technique with two actively synchronized lasers and two metallic targets, nanocomposite NiO:Au thin films were fabricated. The fluence of the Nd:YAG laser, which was used for the deposition of Au, determined the volume fraction of Au into the NiO film matrix. This was verified by SEM/EDS measurements as well as by RBS experiments. The latter have shown that there exists a linear dependence of Au amount on the fluence of the Nd:YAG laser for the employed fluence range. Optical absorption measurements have revealed the SPR band of the Au nanoparticles in the 590 nm range. Further, the optical bandgap of NiO decreased monotonically with increasing Au content, which is promising for photo-catalytic and optoelectronic applications. The presence of Au nanoparticles in the NiO thin film proved to have a drastic effect on the application to hydrogen sensing, improving the sensor performance. Furthermore, Au nanoparticles functioned as catalysts that allowed the detection of hydrogen down to concentration levels of a few ppm at operating temperatures as low as 135 $^{o}$C. The NiO:Au nanocomposite with 44.4% Au showed a peak of the hydrogen sensing response around 128 $^{o}$C, which is a first step towards gas selectivity by a sensor array.

*Acknowledgements*

*The authors would like to acknowledge the financial support of the Hellenic General Secretariat for Research and Technology through a bilateral Greek-Slovak Research Agreement (2006-08) as well as a partial support from the "Nano-structured organic-inorganic hybrid materials – synthesis, diagnostics and properties'' program No.2005ΣE01330081 of TPCI/NHRF as a "Centre of Excellence", 2005. One of the authors (I.F.) would also like to thank the TPCI/NHRF for the financial support in the frame of a 2 years scholarship.*
*They also want to thank Drs. A. Lagoyannis and S. Harisssopoulos, from the Institute of Nuclear Physics, NCSR 'Demokritos', Athens, Greece, for the recording of the RBS spectra at the tandem ion accelerator.*

**Figure captions**

Figure 1. Experimental setup for two-laser, two target pulsed laser deposition.

Figure 2. SEM image (a) and EDS spectra from the "film background" (b) and from a droplet (c) of a NiO thin-film reference sample.

Figure 3. EDS spectra from the "film background" of NiO:Au nanocomposites deposited with 1.3 J/cm$^2$ (a) and 4.3 J/cm$^2$ (b) fluence of the Au ablating laser. (c) SEM image and (d) EDS spectrum from the droplet indicated in (c) for the NiO:Au nanocomposite film deposited with 4.3 J/cm$^2$.

Figure 4. (a) RBS spectrum of the NiO:Au nanocomposite deposited with 1.3 J/cm$^2$ laser fluence, (b) plot of Au concentration vs. laser fluence.

Figure 5. (a) Absorbance spectra for NiO:Au nanocomposites, the reference NiO thin film, and quartz, (b) Tauc plot for reference NiO and NiO:Au nanocomposites.

Figure 6. Hydrogen sensing in air under static gas conditions, obtained with (a) reference NiO sample and (b) NiO:Au nanocomposite with 44.4wt% Au concentration (4.3 J/cm$^2$ laser fluence). The chamber was purged with air ("air flow") between successive H$_2$ cycles, in order to remove the static H$_2$/air mixture and prepare for the next measurement.

Figure 7. Summary of sensing results for all temperatures and hydrogen concentrations in air, obtained (a) with the reference NiO sample, (b) with NiO:Au nanocomposite with 15.6wt% Au concentration (1.3 J/cm$^2$ laser fluence), (c) with NiO:Au nanocomposite with 23.5wt% Au concentration (2.4 J/cm$^2$ laser fluence), and (d) with NiO:Au nanocomposite with 44.4wt% Au concentration (4.3 J/cm$^2$ laser fluence).

Figure 8: Hydrogen sensing in air under dynamic flow conditions and hydrogen dilution for low concentrations, obtained with the NiO:Au nanocomposite with 44.4wt% Au concentration (4.3 J/cm$^2$ laser fluence).

**Table caption**

Table 1: Au concentration and thickness for the reference thin-film NiO sample and nanocomposite NiO:Au samples grown with different laser fluences for the Au ablating laser.

**Table 1**

**Table 1 RBS and electrical results**

| Sample | NiO | NiO:Au 1.3 J/cm$^2$ | NiO:Au 2.4 J/cm$^2$ | NiO:Au 4.3 J/cm$^2$ |
|---|---|---|---|---|
| **Au** (wt%) | - | 15.6 ±0.8% | 23.5 ±1.5% | 44.4 ±2.2% |
| **Thickness (nm)** | 290 | 500 | 590 | 600 |

**Figure 1**

Fig.1

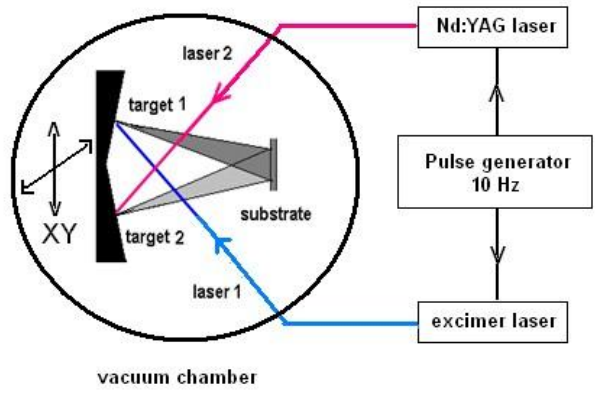

**Figure 2**

Fig. 2

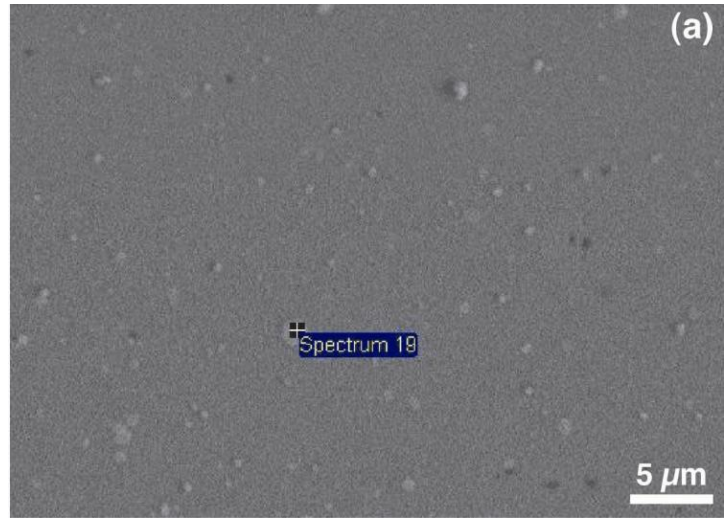

(a)

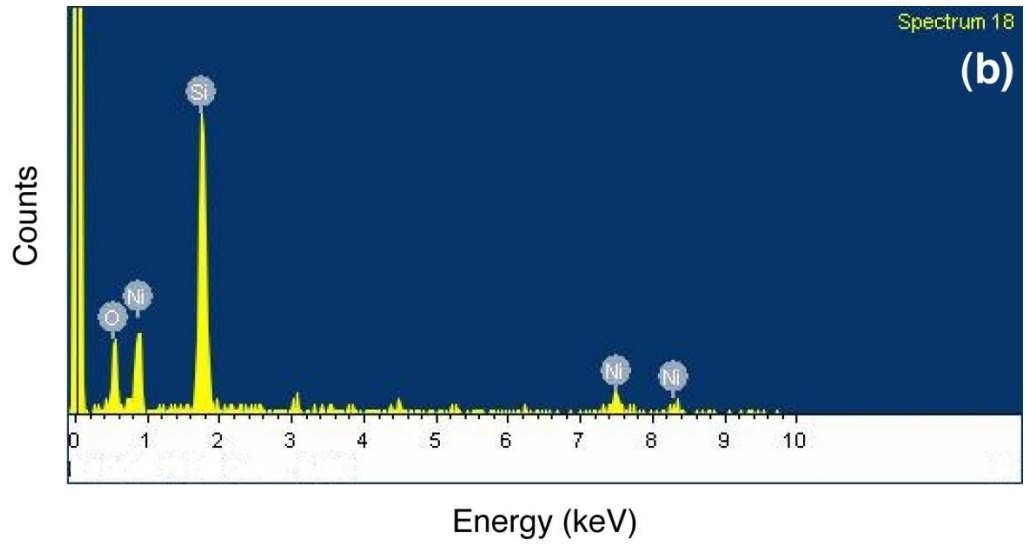

(b)

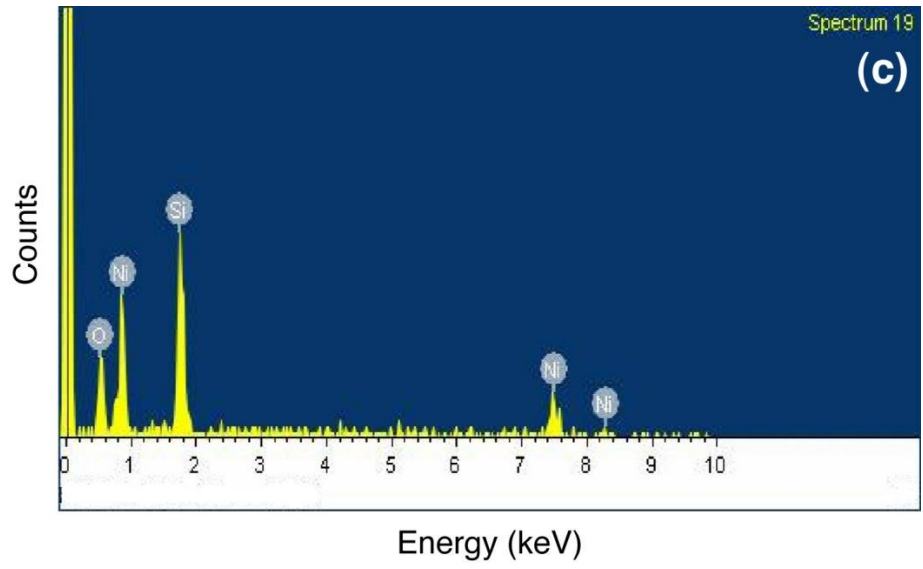

(c)

**Figure 3**

Fig. 3

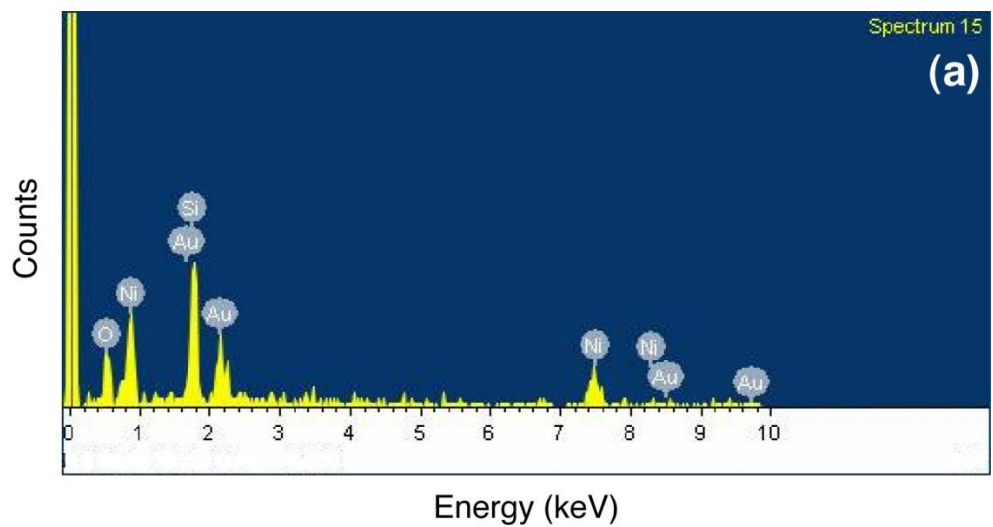

(a)

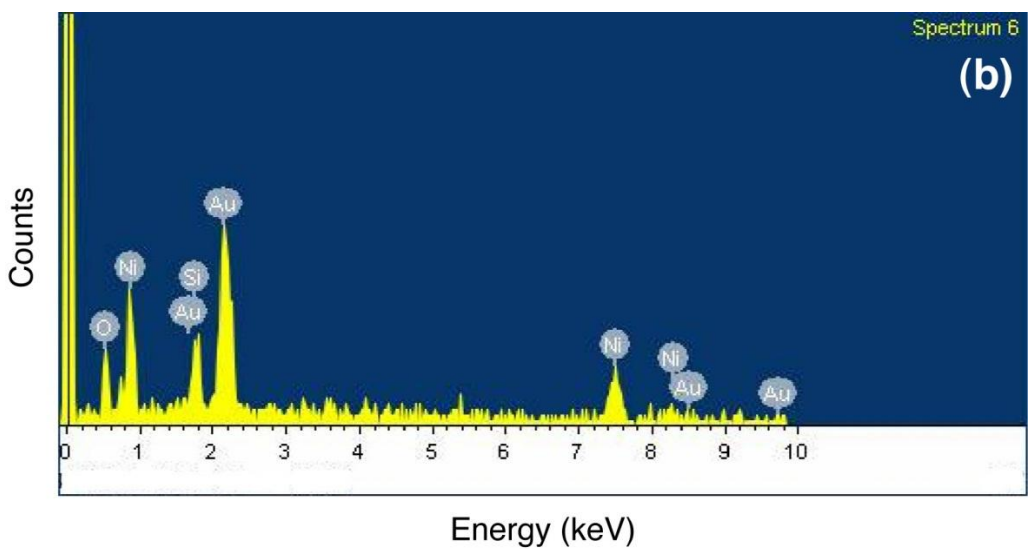

(b)

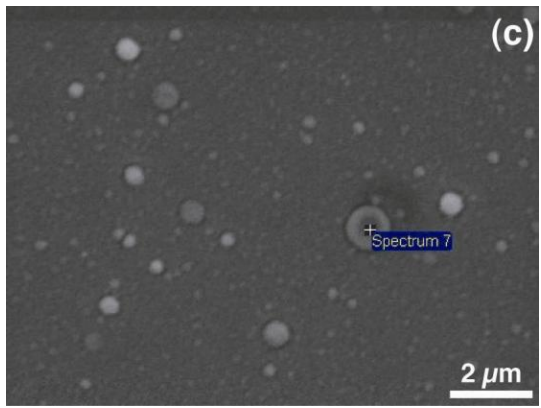

(c)

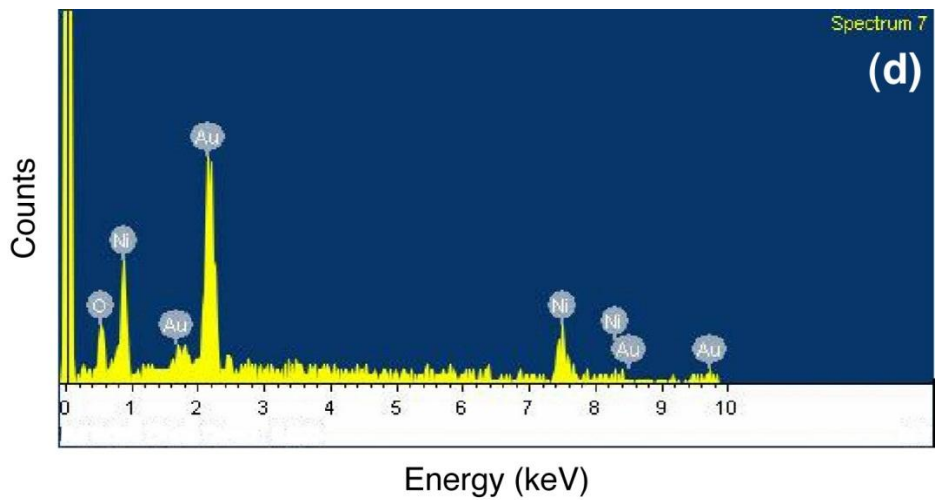

(d)

**Figure 4**

Fig. 4ab RBS

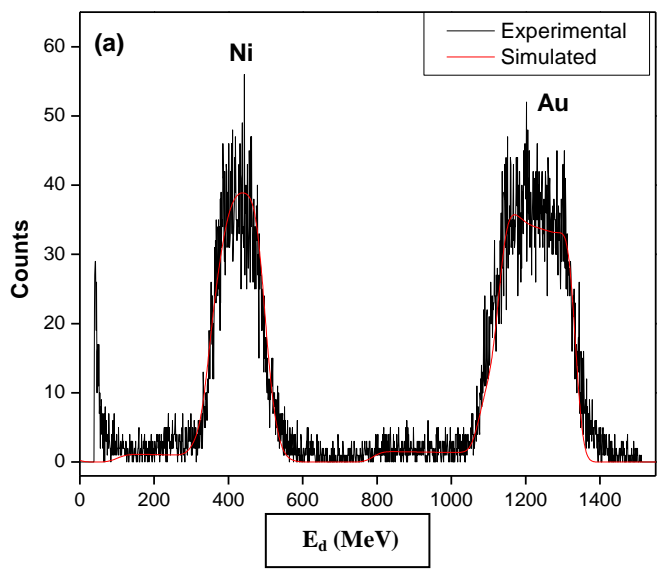

(a)

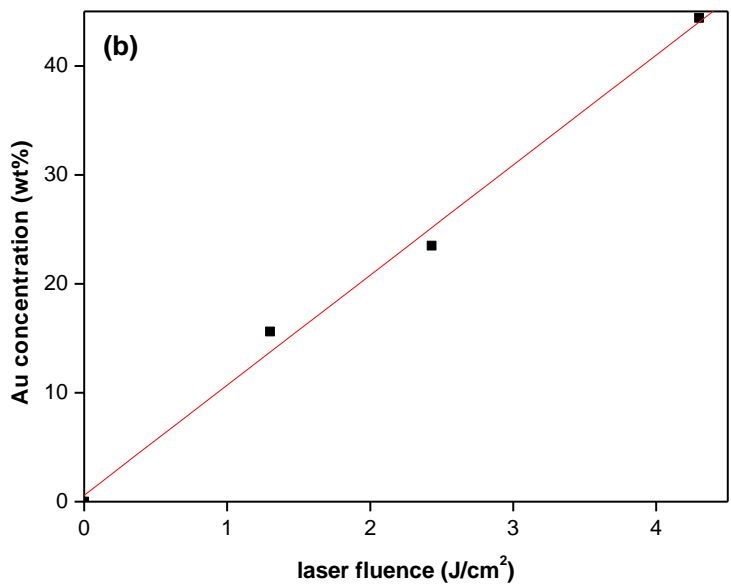

(b)

**Figure 5**

Fig. 5ab OPTICAL

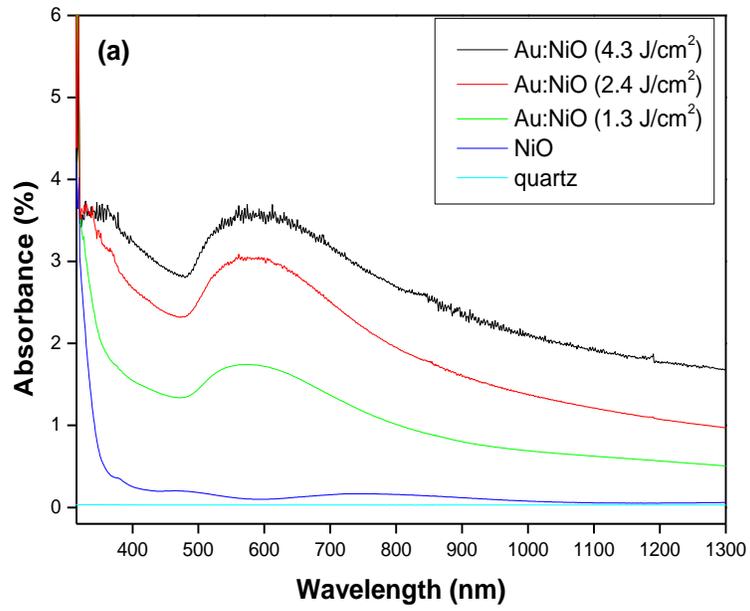

(a)

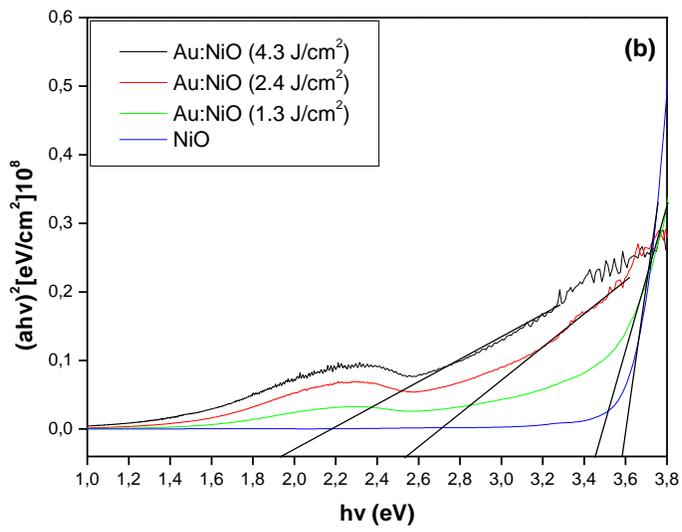

(b)

Fig. 6a

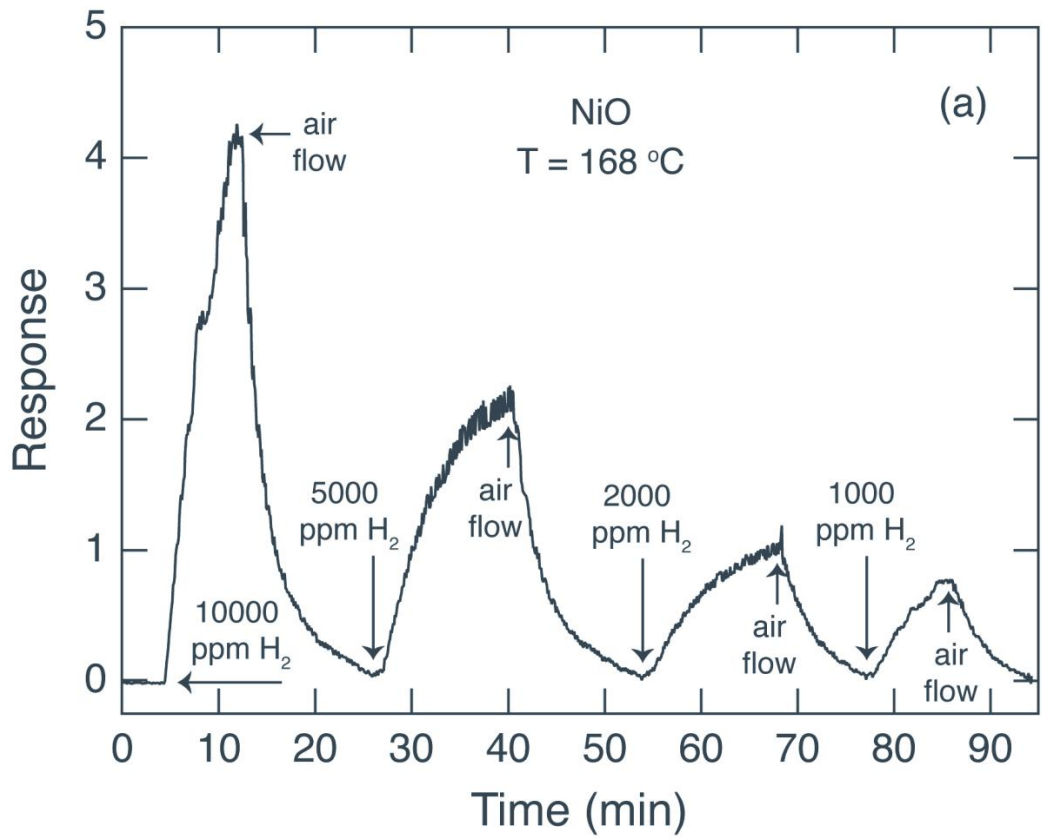

Fig. 6b

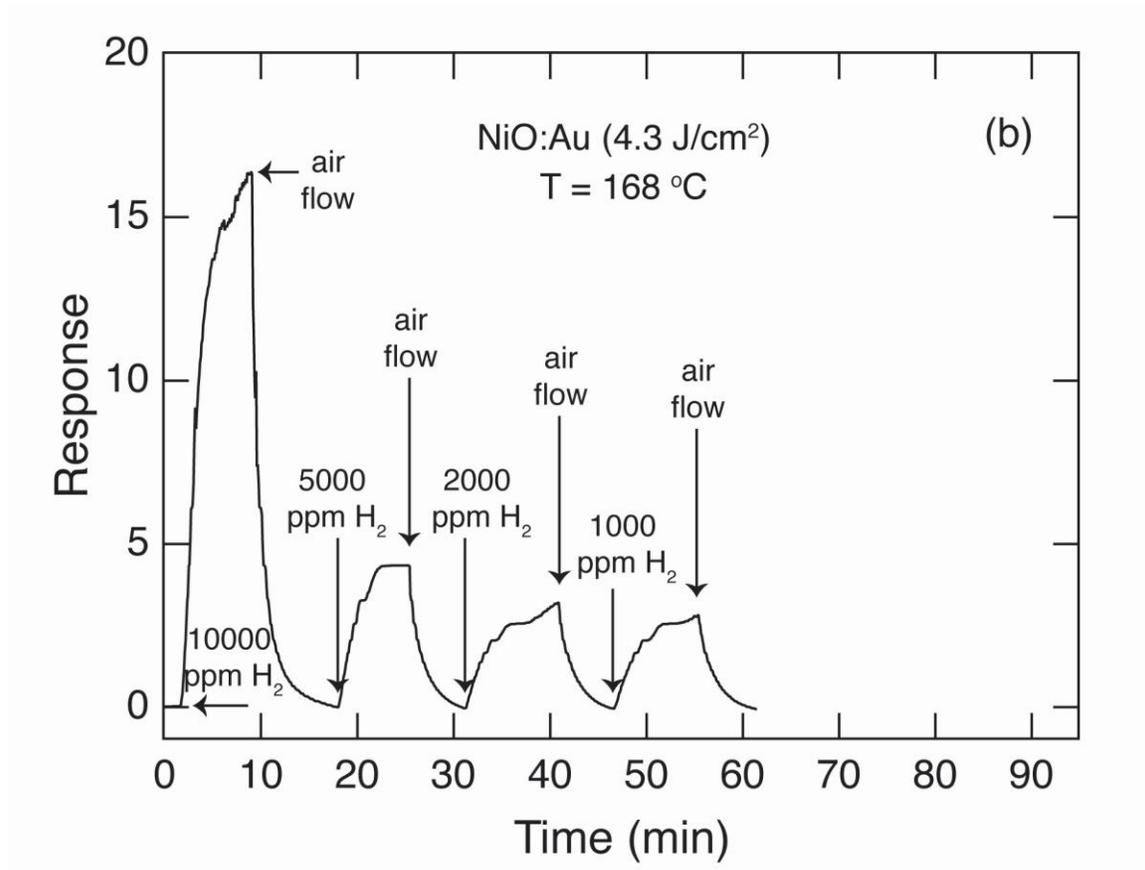

Fig. 7

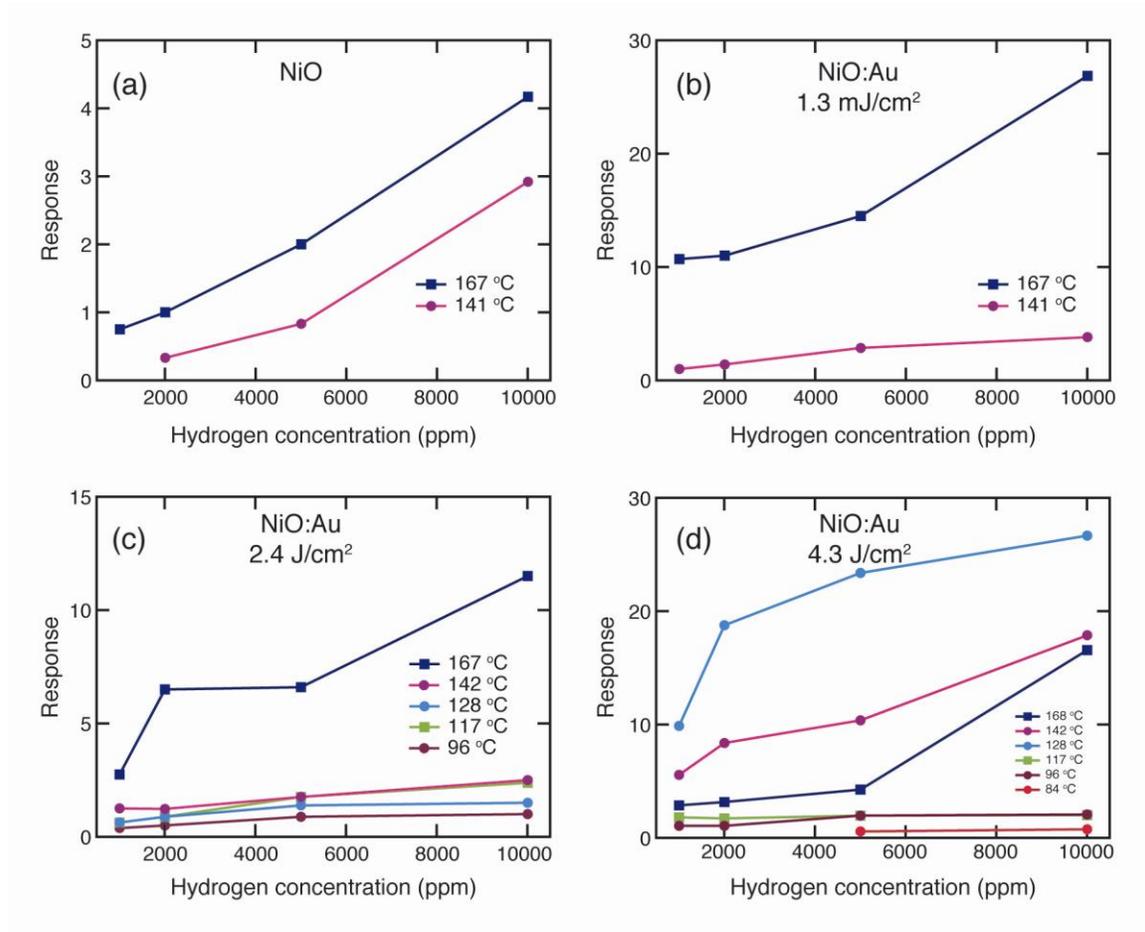

Fig. 8

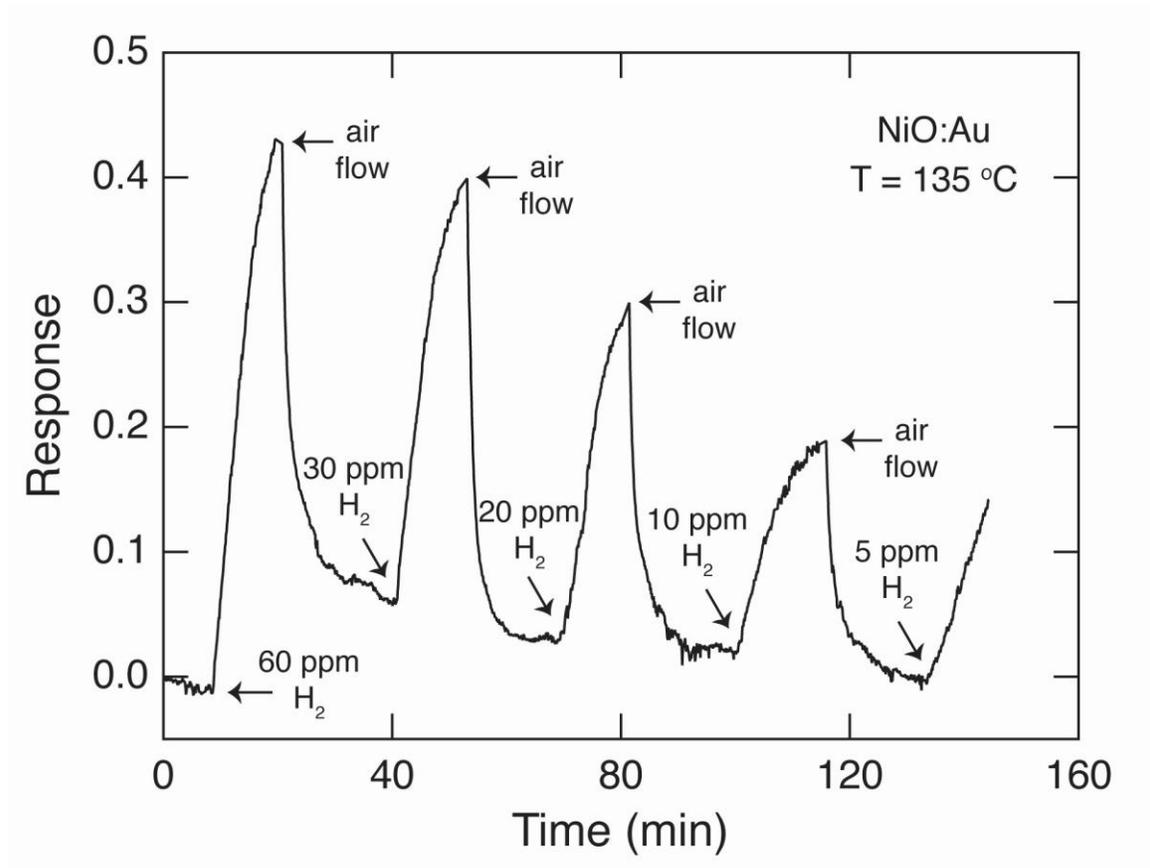